\documentclass[preprint]{aastex}
\shorttitle{Prominence-horn dynamics}

\shortauthors{Wang et al.}

\usepackage{graphicx}
\usepackage{cite}

\begin{document}
\title{Dynamics of a prominence-horn structure during its evaporation in the solar corona}

\author{Bing Wang\altaffilmark{1}, Yao Chen\altaffilmark{1},
Jie Fu\altaffilmark{1, 2}, Bo Li         \altaffilmark{1}, Xing Li
\altaffilmark{3}, and Wei Liu\altaffilmark{4}}

\altaffiltext{1}{Shandong Provincial Key Laboratory of Optical
Astronomy and Solar-Terrestrial Environment, and Institute of
Space Sciences, Shandong University, Weihai 264209, China;
yaochen@sdu.edu.cn}

\altaffiltext{2}{Department of Physics and Astronomy, University
of British Columbia, Vancouver, BC V6T 1Z1, Canada}

\altaffiltext{3}{Department of Physics, Aberystwyth University,
Aberystwyth, Ceredigion, SY23 3BZ, UK}

\altaffiltext{4}{Stanford-Lockheed Institute for Space Research,
Stanford University, Stanford, CA 94305, USA}

\begin{abstract}
The physical connection among and formation mechanisms of various
components of the prominence-horn cavity system remain elusive.
Here we present observations of such a system, focusing on a
section of the prominence that rises and separates gradually from
the main body. This forms a configuration sufficiently simple to
yield clues to the above issues. It is characterized by embedding
horns, oscillations, and a gradual disappearance of the separated
material. The prominence-horn structure exhibits a large amplitude
longitudinal oscillation with a period of $\sim$150 minutes and an
amplitude of $\sim$30 Mm along the trajectory defined by the
concave horn structure. The horns also experience a simultaneous
transverse oscillation with a much smaller amplitude ($\sim$3 Mm)
and shorter period ($\sim$10-15 minutes), likely representative of
a global mode of the large-scale magnetic structure. The gradual
disappearance of the structure indicates that the horn, an
observational manifestation of the field-aligned transition region
separating the cool and dense prominence from the hot and tenuous
corona, is formed due to the heating and diluting process of the
central prominence mass, while most previous studies suggest that
it is the opposite process, i.e., the cooling and condensation of
coronal plasmas, to form the horn. This study also demonstrates
how the prominence transports magnetic flux to the upper corona, a
process essential for the gradual build-up of pre-eruption
magnetic energy.
\end{abstract}


\section{Introduction}
Prominences, cool and dense materials in the solar atmosphere, are
frequently accompanied by coronal cavities and  horn-like
structures. This is especially true for high-latitude quiescent
prominences \citep[e.g.][]{2011A&A...533L...1R,
2012ApJ...758L..37B, 2013ApJ...770...35S}. Coronal cavities are an
elliptical and relatively dark region observed in EUV at typical
coronal temperatures (1 - 2 MK). Horns are bright and narrow
concave-upward structures emanating from prominences and extending
into cavities, best visible in the 171~\,\AA{} passband of the
Atmospheric Imaging Assembly \citep[AIA;][]{2012SoPh..275...17L}
on board the Solar Dynamic Observatory
\citep[SDO;][]{2012SoPh..275....3P}. The prominence-cavity system
is in general believed to be representative of a flux rope
structure consisting of twisted magnetic field lines, which is one
major progenitor of solar coronal mass ejections (CMEs; Fan 2001;
Fan \& Gibson 2003; Gibson \& Fan 2006; Manchester et al. 2004;
Schmit et al. 2009, see Luna et al. (2012) and Schmit et al.
(2013) for an alternative scenario with a sheared arcade and
overlying unsheared field for the prominence-horn-cavity system).
It remains elusive as to how the magnetic energy and flux are
gradually transported to and accumulated in the corona, thereby
leading to eruptions.

It is generally believed that prominences are suspended at
magnetic dips of a larger magnetic field configuration and
supported by the magnetic tension force. On the other hand, horns
connect the cool and dense prominence to the hot and tenuous
corona, representing a field-aligned prominence-corona transition
region (e.g., Luna et al. 2012). The morphology of horns indicates
that they are field-aligned structures, and can therefore be used
to infer the magnetic configuration of the prominence-cavity
system. A series of studies suggest that horns are a part of the
formation process of prominences from cooling and condensing
coronal plasmas \citep{2012ApJ...745L..21L, 2012ApJ...758L..37B,
2013ApJ...770...35S, 2013ApJ...779..156S, 2012ApJ...746...30L}.
This has been supported by a detailed examination of spatial and
temporal correlations between emission intensities observed in
various AIA passbands, for a few sets of horn structures, and
supported by radiative magnetohydrodynamic (MHD) simulations
(e.g., Luna et al. 2012, Schmit et al. 2013 and Xia et al. 2014).
A different scenario of horn formation has been suggested by
\citet{2012ApJ...758...60F} , who proposed that horns are the
current concentration and potential or ongoing reconnection (and
thus heated) site around magnetic separatrices of a hyperbolic
flux tube. This second scenario is based on the three-dimensional
MHD simulations of coronal cavities. However, so far it has no
observational support to the best of our knowledge. Here we will
present a third scenario to explain the horn formation.

Prominence-filament oscillations have been well observed for
nearly a century with ground- and space-based instruments 
\citep[e.g.][]{1930MNRAS..91..239D, 1935MNRAS..95..650N,
1960PASP...72..357M, 2003ApJ...584L.103J, 2004ApJ...608.1124O,
2014ApJ...785...79L}. They can be classified into longitudinal and
transverse oscillations according to the orientation of
oscillations relative to filament spines (see
\citet{2014ApJ...795..130S} for a recent study and more
references), and into small  (a few km~s$^{-1}$) and large
($>$10~km~s$^{-1}$) amplitude oscillations. Small amplitude
oscillations are common in the solar atmosphere, whereas large
amplitude ones are relatively rare. Prominence oscillations are
important because they carry valuable information on the
underlying magnetic field structure, which is difficult to measure
with present techniques. This line of studies forms a research
branch of solar physics, dubbed prominence seismology
\citep[e.g.][]{1966ZA.....63...78H, 2002SoPh..206...45O,
2009SSRv..149..283T, 2012LRSP....9....2A,
 2014ApJ...785...79L}.

Among various types of prominence oscillations, the most relevant
to our study is the large-amplitude longitudinal oscillations
(LALOs), first reported by \citet{2003ApJ...584L.103J}. According
to the latest numerical studies \citep{2012ApJ...750L...1L,
2016ApJ...817..157L} the most likely restoring force for LALOs is
the field-aligned component of solar gravity, in much the same way
that pendulums operate. Indeed, the observed periods seem to be
consistent with the pendulum theory. Here in our event not only an
LALO itself can be clearly observed, but also we can infer the
associated magnetic curvature with the accompanying horn
structure. This provides a nice opportunity to test the pendulum
theory of LALOs.

Regarding the intimate prominence-horn connection and the rich
varieties of prominence oscillations, it is natural to ask whether
the horns can also manifest oscillations. The numerical study by
Luna et al. (2016) shows that besides the prominence LALO, the
large-scale magnetic structure may also support a periodic
transverse motion that is interpreted as a global fast normal
mode. The horns with their relatively-long field-aligned spatial
extension serve as a good magnetic tracer to reveal the existence
of such modes, as supported by our novel observation of horn
dynamics.

\section{Observational data and event overview}
The prominence examined here is a typical quiescent one, observed
by both the AIA/SDO and the H$_\alpha$ images provided by the
Global Oscillations Network Group (GONG) project. It appeared in
the field of view (FOV) at the northeast limb ($\sim$45$^\circ$ in
latitude) at 22:00 UT on July 7, 2011. According to Figure 1a-1c
and the accompanying animation, the prominence pillar emits at
H$_\alpha$ ($\sim 10^4$ K) and 304~\AA{} ($\sim 0.08$ MK), and
manifests as a vertical absorption feature at 171~\AA{} ($\sim
0.8$ MK). The southern side of the H$_\alpha$ pillar is connected
to the surface by a number of arching threads which are
also observable in the hotter 304-171~\AA{} passbands. Within
these threads persistent bi-directional mass flows are observed,
indicative of a dynamical equilibrium undergoing significant mass
exchange. This is consistent with earlier studies on the mass
supply and drainage processes of solar prominences
\citep[e.g.][]{1998Nat...396..400, 2011Natur.472..197B,
2012ApJ...758L..37B, 2012ApJ...745L..21L}.

From the accompanying animation we see that the upper part of the
prominence presents co-incident wavy motion in both the H$_\alpha$
and 304~\AA{} passbands, yet the H$_\alpha$ images have in general
poorer quality with less details than the 304~\AA{} data. In
addition, some structures appear above the prominence pillar at
304~\AA{} while invisible at H$_\alpha$. This is likely due to the
progressively more rarefied and hotter condition of the
emitting materials associated with the rising flux tube higher in
the corona. Dynamics of these structures observed by AIA at 304
and 171~\AA{} are the focus of our study.

It can be seen that the prominence has a similar morphology in the
171 \AA{} emission (Figure 1b), on top of which we observe a group
of concave-upward structures that have been dubbed prominence
``horns'' \citep[e.g.][]{2012ApJ...758L..37B} or ``U-shape
structures'' \citep{2011A&A...533L...1R}. Consistent with previous
studies, the horns emanate from the corresponding prominence and
extend upward into the corona. In Figure 1d, we present the AIA
211 \AA{} image recorded at 03:28 UT on July 10, which shows the
associated cavity on top of the prominence-horn structure.
Although the cavity is only clearly visible after 02:00 UT on July
10, the overall prominence-horn morphology does not change much.
We therefore suggest that the underlying structure, presumably a
flux rope, has been present during the earlier period of interest.

The prominence-cavity structure starts to rise gradually at 08:00
UT and erupts at 11:30 UT on July 10, leading to a CME propagating
at a linear speed of $\sim  $427~km~s$^{-1}$ according to the
Large Angle Spectroscopic Coronagraph
\citep[LASCO;][]{1995SoPh..162..357B} C2 data. The pre-eruption
prominence is observed on the disk by the Solar TErrestrial
RElations Observatory \citep[STEREO;][]{2008SSRv..136....5K} B,
yet it is very faint, possibly due to its high altitude
as indicated by the simultaneous AIA images and its relatively
weak emission in comparison with other disk features. The eruption
results in a large-scale arcade structure in the corona according
to the STEREO-B 304 \AA{} data (not shown here), with flare-like
ribbons sweeping across a region $\sim$50$^\circ$ long in
longitude and $\sim$400 Mm (55 arcsecs) wide. This suggests that
the pre-eruption prominence-cavity system is a well-developed,
hemispheric-scale magnetic structure.

Here we focus our study on the prominence-horn dynamics observed
one day before the eruption. Of particular interest is the
dynamical evolution when a small upper section of the prominence
rises and separates from the underlying main prominence structure.
At $\sim$19:00 UT on July 8, this section, together with the
embedded horns, appears mostly isolated from the main structure
(See the region S1 in Figure 1e-1f and the animation). We
speculate that the separation and the accompanying large scale
motion are possibly due to reconnections, prominence structural
changes or a mass drainage (or unloading) process that destroys
the original force balance around the top of the prominence.

This upper section then ascends gradually into the overlying
structure which has been suggested to be a flux rope and
manifested later as a corona cavity (see Figure 1d). The
suspending prominence-horn structure exhibits a large-amplitude
oscillation while slowly fading from the FOV and disappearing
completely around 20:50 UT. We concentrate on the prominence-horn
dynamics from the upper prominence section starts to ascend
($\sim$17:00 UT) to its disappearance. While the structure of
interest is also visible at 131 \AA{} through its low-temperature
response ($\sim$0.5 MK), the emission is too weak to reveal useful
information.

Because the main prominence body is structurally complex
and highly dynamic, only the significant separation of the small
upper section from the main body as well as its projection onto
the relatively dark coronal background allows us to clearly
isolate this portion of the prominence for further analysis. 

\section{Main results}
As is evident from the animation accompanying Figure 1, the upper
part of the prominence exhibits a large-scale motion along a
curved trajectory that basically follows the curvature of the
embedding horns.

The motion can be seen from Figure 2 which presents the
distance-time map of S1, in which the 304~\AA{} intensities have
been averaged along the vertical direction. We can see that the
prominence starts to move northwestward at $\sim$16:30 UT, and
reaches the end at $\sim$17:10 UT, after traversing a distance of
$\sim$15 Mm. Then the prominence reverses its direction and moves
southeastward before moving northwestward again. During this
process, the prominence disintegrates into several pieces with
each piece accompanied by thread-like horn structures. The bulk
collective motion cannot be clearly defined after 19:00 UT at
304~\AA{} and 19:30 UT at 171~\AA{}. As mentioned, this
large-scale prominence motion is basically along the associated
field-aligned horn features, and therefore in accordance with the
observational manifestation of LALOs, albeit the motion does not
even finish a complete cycle. The period of our LALO is estimated
to be $\sim$150 minutes and the amplitude is $\sim$30 Mm,
consistent with earlier LALO reports
\citep[e.g.][]{2003ApJ...584L.103J, 2006ApJ...644.1273J,
2007A&A...471..295V, 2014ApJ...795..130S, 2014ApJ...785...79L}.

The restoring force of the LALO has been suggested to be the
field-aligned component of the gravity, similar to a pendulum with
a length given by the curvature radius of the supporting magnetic
structure. The presence of accompanying horns allows us to measure
the associated magnetic curvature radius. This provides a nice
opportunity to test the pendulum theory of LALOs. To do this, we
select several clear horn structures during the LALO and measure
their curvature radii. The selected horns are plotted with dashed
curves in Figure 3, and are fitted with circular arcs. The
curvature radii thus found are 73, 70, and 66 Mm for the three
horns, respectively. The corresponding average is $\sim$70 Mm,
leading to an expected pendulum period of $\sim$1h. This is
considerably less than the measured value of $\sim$2.5 h. Possible
factors leading to this discrepancy will be discussed in our final
section.

To explore the horn dynamics during the above process, in Figure 4
we plot distance-time maps along three slices across the horns (S2
- S4, see Figure 3). We see that the eastern part of the horn H1
manifests a clear quasi-periodic oscillation (See also the online
animation). The oscillation is mostly perpendicular to the horn,
which is suggested to delineate a bundle of magnetic field
lines, and thus representative of a transverse mode. It is
clearly observed from 17:00 - 18:00 UT, lasting for several
periods. The period is $\sim$ 10 - 15 minutes and the amplitude is
$\sim$ 3 Mm. Both parameters, much smaller than those of the LALO,
are comparable to earlier results of small-amplitude transverse
prominence oscillations (SATPOs). Since horns emanate from the
prominence, it is natural to surmise that the horn oscillations
are associated with the SATPO. From the S3 map, which crosses the
middle part of the prominence, there seems to exist a weak
oscillation with a smaller amplitude and a similar period, yet the
feature is diffuse and our statement is not conclusive. An
inspection of the accompanying animation shows that the western
part of two nearby horns also presents some oscillations with
seemingly-opposite phases. However, the features of these
oscillations can be hardly recognized from the distance-time map
along S4 (Figure 4c), due to the rather weak emissivity there. In
other intervals, e.g., after 19:30 UT, signatures of horn
oscillations with comparable periods can be seen from the first
two panels of Figure 4, indicating that the horn motion
investigated here possibly represents a persistent global
transverse mode supported by the large-scale magnetic structure.

Now we examine the emission features that might reveal some clues
to the formation mechanism of horns. We first plot the 304 and 171
\AA{} emission intensity variation along the 3 horns (see Figure
3) in Figure 5a-5c. The intensities have been averaged over 5
pixels transversely. We see that the 171~\AA{} intensity tends to
reach a high level around the 304~\AA{} intensity maximum. Yet,
the details differ for different horns. For H1, the 171~\AA{}
emission reaches a local minimum, anti-correlated with the
304~\AA{} intensity profile, while for H2 and H3 the 171~\AA{}
emissions reach a plateau of their intensities around the
304~\AA{} maxima. These observations are not inconsistent with
those modelled by Luna et al. (2012) with a simplified
one-dimensional thermal non-equilibrium model of coronal
condensation and prominence formation. They showed that the horns
at 171~\AA{} are characterized by brightenings on both sides of a
central 304~\AA{}-emitting mass. From the central part towards
ends along the horns, both 171 and 304 \AA{} emissions decline
gradually. Note that the 171 intensity always declines slower and
occupies a broader emitting region than its 304 counterpart.

Overall, the 171 and 304 \AA{} emissions are strongly correlated,
as clearly seen from their correlated variation trends shown in
the four panels of Figure 5. The last panel (5d) presents their
intensity curves, given by the corresponding total normalized
intensities in S1. We see that the two curves reach the maximum
intensity at 18:47 (304~\AA{}) and 18:57 (171~\AA{}) UT,
respectively, before gradually fading out. It seems that the
normalized 171 emission maximum is somehow delayed by $\sim$ 10
minutes. A similar delay is also indicated by the later knee-like
feature around 19:10-19:20 UT. The significance of this delay
shall be further investigated and testified with more events.

The observed close correlation of the 171 and 304 emission
intensities has been demonstrated previously by
\citet{2013ApJ...779..156S} who suggested that horns are formed
due to the cooling-condensation process of coronal plasmas,
associated with the prominence formation. In our case the
dynamical and morphological evolution of horns is observed during
the gradual disappearance of a prominence section. This strongly
indicates that the opposite process, referred to as the prominence
\emph{evaporation}, might take place here with the horns being
formed by the heating and diluting process of prominence plasmas.

As clearly seen from the animation accompanying Figure 1, the
ascending part of the prominence gradually disappears into the
upper coronal magnetic structure that is observed one-day later as
a cavity (presumable a flux rope structure, see Figure 1d). During
this process, the structure disintegrates into several parts which
are initially tied together through the prominence mass. Each part
carries a horn structure that is representative of a bundle of
field lines. Thus, we suggest that the observed prominence ascent
and evaporation actually represent a process of the magnetic flux
(as well as mass and magnetic energy) transport from the
large-scale prominence to the upper cavity structure. It is
observable here because of the presence of the field-tracing
horns. This process certainly contributes to the gradual flux
accumulation and energy build-up in the pre-eruption corona, and
may play a role in the onset of the one-day-later CME.

\section{Summary and discussion.}
Various dynamical processes of a prominence-horn structure are
investigated to provide insights into the prominence-horn cavity
connection and accompanying oscillations. The event is observed
during a gradual rising and a likely-evaporation process of a
prominence section that separates from the main body. The
prominence section of interest, with its embedded horns,
experiences a large amplitude oscillation with a period of
$\sim$2.5 h and an amplitude of $\sim$ 30 Mm along the trajectory
defined by the concave-outward horn structure. The simultaneous
observation of this LALO and the associated horn structure offers
a nice opportunity to test the pendulum theory of the LALO and it
is found that the predicted period is significantly shorter than
observed.

The mechanisms underlying the possible evaporation could not be
determined with available data. Presumptions include the
field-aligned thermal conduction from the hot corona plasmas
towards the cool prominence plasmas, changes of the amount and
spatial distribution of heating along the flux tube with horns
(see, e.g., Luna et al., 2012), and/or waves- or turbulence-
related heating processes within or outside the cool prominence
section. In addition, the rising of the prominence-horn structure
into the higher corona may cause the field line to expand with a
shallower dip geometry, and carry the cool dense material into a
less dense environment. This may also contribute to
the draining and evaporation of the prominence material.

As mentioned in the introduction, latest numerical studies (Luna
et al. 2012, 2016) suggest that the LALO is mainly controlled by
the gravitational force, and its damping is directly related to
the rate of mass accretion that is induced by
footpoint-heating-caused chromospheric evaporation and further
condensation. Our LALO event takes place during the gradual
disappearance of cool materials. Therefore, the mass accretion is
not important here. We propose several factors that may contribute
to the above-mentioned discrepancy. First of all, since in our
case the prominence moves higher to a progressively more
rarefied environment with emitting materials contained by a
rising flux tube of weaker magnetic field, the relative importance
of the gravity and the thermal coupling process in
governing the plasma dynamics may differ from their roles in the
static flux tubes investigated numerically. This, demanding a
further study, may partially account for the reported discrepancy.

Other possible factors include (1) the projection effect of the
horn structure that leads to an underestimate of the curvature
radius and thus the expected pendulum period, (2) the overall
deviation of the magnetic structure away from the radial direction
thus making only part of the solar gravity the restoring force,
also resulting in an underestimate of the predicted period, and
(3) the thermal and magnetic coupling between the rising part and
the lower main body may enhance the restoring force and thus
reduces the period. Among these factors, the effect of the second
one can be roughly estimated here. From our data, the horn
structure, along which the LALO takes place, is at an angle of
$\leq$20 degrees away from the radial direction, resulting in a
slightly longer period ($\sim$1.1h) than predicted using the total
gravity, yet still much shorter than observed ($\sim$2.5h).

In addition to the LALO, the horns also undergo a transverse
oscillation with a smaller amplitude ($\sim$3 Mm) and shorter
period ($\sim$10-15 minutes). This oscillation seems to be carried
by several horns across an extended region and lasts for a
relatively long time. This finding is consistent with the latest
numerical study of prominence oscillations by Luna et al. (2012, 2016), 
which reveals the persistent existence of a global fast collective
mode of the magnetic structure that supports an oscillating
prominence. Their results are comparable to those reported here
for the small-amplitude transverse horn oscillations, indicating
that the oscillation is likely the fast collective mode supported
by the overall magnetic structure. More studies are required to
further explore the seismological potential of this kind of
oscillations.

As mentioned in Section 1, currently available scenarios to
explain the horn formation include coronal condensation
\citep{2012ApJ...758L..37B, 2013ApJ...770...35S,
2013ApJ...779..156S}, and the current-sheet concentration surface
of a hyperbolic flux tube \citep{2012ApJ...758...60F}.
What we report here shows that the horn structure emanates from
the prominence and gradually fades away. This strongly indicates
that the horn formation may also be attributed to the prominence
evaporation, likely involving some heating and diluting process of
the central prominence mass. This can be regarded as a scenario
opposite to the present cooling-condensation scenario of horn
formation. In the corona, it may go both ways, heating or cooling,
both with a transition between cool and hot mass.

Using the horn structure as tracers of the underlying magnetic
structure, our study also illustrates how the prominence can
transport magnetic flux to the overlying cavity-flux rope
structure, and thus to increase the magnetic energy there. This
may be a fundamental process essential for the gradual build-up of
magnetic energy in the pre-eruption corona, especially in the
cavity region overlying a prominence structure.

\acknowledgments

We gratefully acknowledge the usage of data from the SDO, STEREO,
SOHO spacecraft and the ground-based GONG project. This work was
supported by grants NNSFC 41331068, 41274175 (YC), 41274176,
41474149 (BL), and NSBRSF 2012CB825601.


\begin{thebibliography}{}
\expandafter\ifx\csname natexlab\endcsname\relax\def\natexlab#1{#1}\fi

\bibitem[{{Arregui} {et~al.}(2012){Arregui}, {Oliver}, \&
  {Ballester}}]{2012LRSP....9....2A}
{Arregui}, I., {Oliver}, R., \& {Ballester}, J.~L. 2012, Living Reviews in
  Solar Physics, 9, doi:10.12942/lrsp-2012-2

\bibitem[{{Berger}(2014)}]{2014IAUS..300...15B}
{Berger}, T. 2014, in IAU Symposium, Vol. 300, Nature of Prominences and their
  Role in Space Weather, ed. B.~{Schmieder}, J.-M. {Malherbe}, \& S.~T. {Wu},
  15--29

\bibitem[{{Berger} {et~al.}(2011){Berger}, {Testa}, {Hillier}, {Boerner},
  {Low}, {Shibata}, {Schrijver}, {Tarbell}, \& {Title}}]{2011Natur.472..197B}
{Berger}, T., {Testa}, P., {Hillier}, A., {et~al.} 2011, \nat, 472, 197

\bibitem[{{Berger} {et~al.}(2012){Berger}, {Liu}, \&
  {Low}}]{2012ApJ...758L..37B}
{Berger}, T.~E., {Liu}, W., \& {Low}, B.~C. 2012, \apjl, 758, L37

\bibitem[{{Brueckner} {et~al.}(1995){Brueckner}, {Howard}, {Koomen},
  {Korendyke}, {Michels}, {Moses}, {Socker}, {Dere}, {Lamy}, {Llebaria},
  {Bout}, {Schwenn}, {Simnett}, {Bedford}, \& {Eyles}}]{1995SoPh..162..357B}
{Brueckner}, G.~E., {Howard}, R.~A., {Koomen}, M.~J., {et~al.} 1995, \solphys,
  162, 357

\bibitem[{{Dyson}(1930)}]{1930MNRAS..91..239D}
{Dyson}, F. 1930, \mnras, 91, 239

\bibitem[Fan(2001)]{2001ApJ...554L.111F} Fan, Y.\ 2001, \apjl, 554, L111

\bibitem[{{Fan}(2012)}]{2012ApJ...758...60F}
{Fan}, Y. 2012, \apj, 758, 60

\bibitem[{{Fan} \& {Gibson}(2003)}]{2003ApJ...589L.105F}
{Fan}, Y., \& {Gibson}, S.~E. 2003, \apjl, 589, L105

\bibitem[Gibson \& Fan(2006)]{2006JGRA..11112103G} Gibson, S.~E.,
 \& Fan, Y.\ 2006, Journal of Geophysical Research (Space Physics), 111, A12103

\bibitem[{{Hyder}(1966)}]{1966ZA.....63...78H}
{Hyder}, C.~L. 1966, \zap, 63, 78

\bibitem[{{Jing} {et~al.}(2003){Jing}, {Lee}, {Spirock}, {Xu}, {Wang}, \&
  {Choe}}]{2003ApJ...584L.103J}
{Jing}, J., {Lee}, J., {Spirock}, T.~J., {et~al.} 2003, \apjl, 584, L103

\bibitem[{{Jing} {et~al.}(2006){Jing}, {Song}, {Abramenko}, {Tan}, \&
  {Wang}}]{2006ApJ...644.1273J}
{Jing}, J., {Song}, H., {Abramenko}, V., {Tan}, C., \& {Wang}, H. 2006, \apj,
  644, 1273

\bibitem[{{Kaiser} {et~al.}(2008){Kaiser}, {Kucera}, {Davila}, {St.~Cyr},
  {Guhathakurta}, \& {Christian}}]{2008SSRv..136....5K}
{Kaiser}, M.~L., {Kucera}, T.~A., {Davila}, J.~M., {et~al.} 2008, \ssr, 136, 5

\bibitem[{{Lemen} {et~al.}(2012){Lemen}, {Title}, {Akin}, {Boerner}, {Chou},
  {Drake}, {Duncan}, {Edwards}, {Friedlaender}, {Heyman}, {Hurlburt}, {Katz},
  {Kushner}, {Levay}, {Lindgren}, {Mathur}, {McFeaters}, {Mitchell}, {Rehse},
  {Schrijver}, {Springer}, {Stern}, {Tarbell}, {Wuelser}, {Wolfson}, {Yanari},
  {Bookbinder}, {Cheimets}, {Caldwell}, {Deluca}, {Gates}, {Golub}, {Park},
  {Podgorski}, {Bush}, {Scherrer}, {Gummin}, {Smith}, {Auker}, {Jerram},
  {Pool}, {Soufli}, {Windt}, {Beardsley}, {Clapp}, {Lang}, \&
  {Waltham}}]{2012SoPh..275...17L}
{Lemen}, J.~R., {Title}, A.~M., {Akin}, D.~J., {et~al.} 2012, \solphys, 275, 17

\bibitem[{{Liu} {et~al.}(2012){Liu}, {Berger}, \& {Low}}]{2012ApJ...745L..21L}
{Liu}, W., {Berger}, T.~E., \& {Low}, B.~C. 2012, \apjl, 745, L21


\bibitem[{{Luna} \& {Karpen}(2012)}]{2012ApJ...750L...1L}
{Luna}, M., \& {Karpen}, J. 2012, \apjl, 750, L1


\bibitem[Luna et al.(2012)]{2012ApJ...746...30L} Luna, M., Karpen, J.~T., \& DeVore, C.~R.\ 2012, \apj, 746, 30

\bibitem[Luna et al.(2014)]{2014ApJ...785...79L} Luna, M., Knizhnik, K., Muglach, K., et al.\ 2014, \apj, 785, 79

\bibitem[{{Luna} {et~al.}(2016){Luna}, {Terradas}, {Khomenko}, {Collados}, \&
  {de Vicente}}]{2016ApJ...817..157L}
{Luna}, M., {Terradas}, J., {Khomenko}, E., {Collados}, M., \& {de Vicente}, A.
  2016, \apj, 817, 157

\bibitem[{{Manchester} {et~al.}(2004){Manchester}, {Gombosi}, {DeZeeuw}, \&
  {Fan}}]{2004ApJ...610..588M}
{Manchester}, IV, W., {Gombosi}, T., {DeZeeuw}, D., \& {Fan}, Y. 2004, \apj,
  610, 588

\bibitem[{{Moreton} \& {Ramsey}(1960)}]{1960PASP...72..357M}
{Moreton}, G.~E., \& {Ramsey}, H.~E. 1960, \pasp, 72, 357

\bibitem[{{Newton}(1935)}]{1935MNRAS..95..650N}
{Newton}, H.~W. 1935, \mnras, 95, 650

\bibitem[{{Okamoto} {et~al.}(2004){Okamoto}, {Nakai}, {Keiyama}, {Narukage},
  {UeNo}, {Kitai}, {Kurokawa}, \& {Shibata}}]{2004ApJ...608.1124O}
{Okamoto}, T.~J., {Nakai}, H., {Keiyama}, A., {et~al.} 2004, \apj, 608, 1124

\bibitem[{{Oliver} \& {Ballester}(2002)}]{2002SoPh..206...45O}
{Oliver}, R., \& {Ballester}, J.~L. 2002, \solphys, 206, 45

\bibitem[{{Pesnell} {et~al.}(2012){Pesnell}, {Thompson}, \&
  {Chamberlin}}]{2012SoPh..275....3P}
{Pesnell}, W.~D., {Thompson}, B.~J., \& {Chamberlin}, P.~C. 2012, \solphys,
  275, 3

\bibitem[{{R{\'e}gnier} {et~al.}(2011){R{\'e}gnier}, {Walsh}, \&
  {Alexander}}]{2011A&A...533L...1R}
{R{\'e}gnier}, S., {Walsh}, R.~W., \& {Alexander}, C.~E. 2011, \aap, 533, L1

\bibitem[{{Schmit} \& {Gibson}(2013)}]{2013ApJ...770...35S}
{Schmit}, D.~J., \& {Gibson}, S. 2013, \apj, 770, 35

\bibitem[{{Schmit} {et~al.}(2013){Schmit}, {Gibson}, {Luna}, {Karpen}, \&
  {Innes}}]{2013ApJ...779..156S}
{Schmit}, D.~J., {Gibson}, S., {Luna}, M., {Karpen}, J., \& {Innes}, D. 2013,
  \apj, 779, 156

\bibitem[{{Schmit} {et~al.}(2009){Schmit}, {Gibson}, {Tomczyk}, {Reeves},
  {Sterling}, {Brooks}, {Williams}, \& {Tripathi}}]{2009ApJ...700L..96S}
{Schmit}, D.~J., {Gibson}, S.~E., {Tomczyk}, S., {et~al.} 2009, \apjl, 700, L96

\bibitem[{{Shen} {et~al.}(2014){Shen}, {Liu}, {Chen}, \&
  {Ichimoto}}]{2014ApJ...795..130S}
{Shen}, Y., {Liu}, Y.~D., {Chen}, P.~F., \& {Ichimoto}, K. 2014, \apj, 795, 130

\bibitem[Tripathi et al.(2009)]{2009SSRv..149..283T} Tripathi, D., Isobe, H., \& Jain, R.\ 2009, \ssr, 149, 283


\bibitem[{{Vr{\v s}nak} {et~al.}(2007){Vr{\v s}nak}, {Veronig}, {Thalmann}, \&
  {{\v Z}ic}}]{2007A&A...471..295V}
{Vr{\v s}nak}, B., {Veronig}, A.~M., {Thalmann}, J.~K., \& {{\v Z}ic}, T. 2007,
  \aap, 471, 295

\bibitem[Xia et al.(2014)]{2014ApJ...792L..38X} Xia, C., Keppens, R., Antolin, P., \& Porth, O.\ 2014, \apjl, 792, L38

\bibitem[Zirker et al. (1998)]{1998Nat...396..400}Zirker, J. B., Engvold, O., \& Martin, S. F.
1998, \nat, 396, 400

\end{thebibliography}

\begin{figure}
\epsscale{1.}
\includegraphics[width=1.0\textwidth]{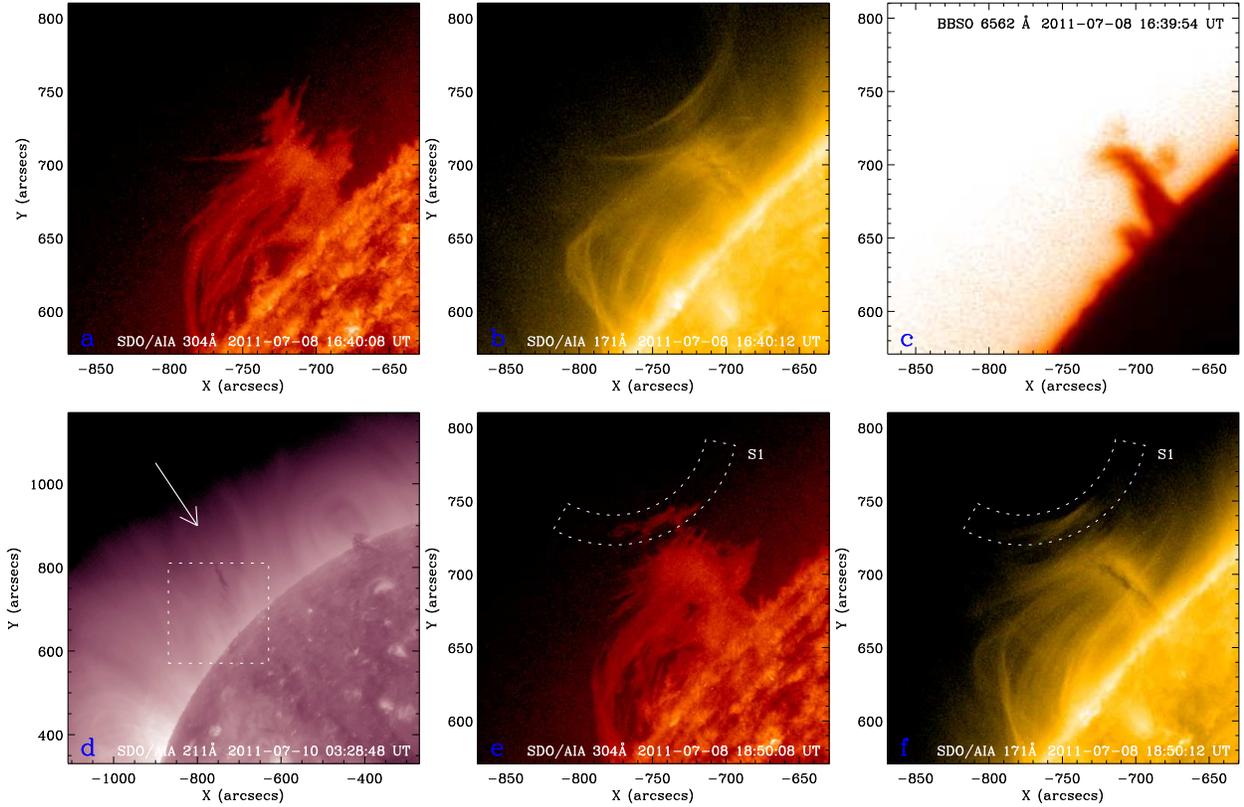}
\caption{Overview of the prominence-horn cavity system and the
associated eruptive features. (a-c) The asymmetric morphology of
the large-scale prominence structure observed in the 304-171 \AA{}
and H$_\alpha$ emissions. (d) The cavity structure at 211 \AA{},
pointed at by a white arrow. The box indicates the FOV of other
panels. (e-f) The separation of the oscillating upper part. The
curved slice (S1) is for the distance-time map shown in Figure 2,
with a starting point located at the ``S1'' end. An animation of
this figure is available (with negative images for better visual
effect).}\label{Fig1}
\end{figure}

\begin{figure}
\epsscale{1.}
\includegraphics[width=2.0\textwidth]{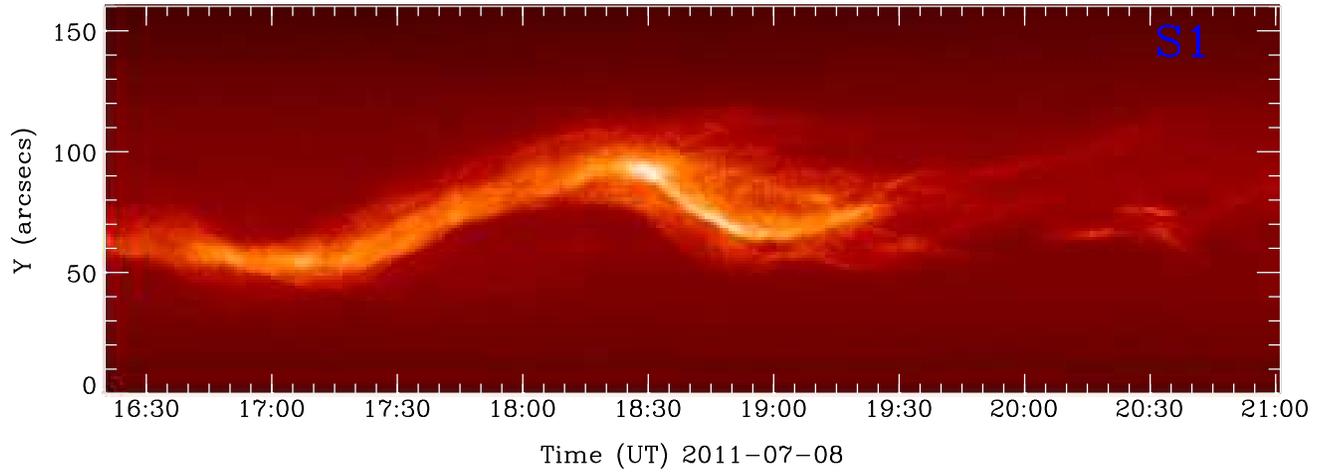}
\caption{Distance-time maps at 304~\AA{} along the slice S1
presented in Figure 1e-1f, to show the prominence LALO motion. The
slice is 268 pixels long and 35 pixels wide.}\label{Fig2}
\end{figure}

\begin{figure}
\epsscale{1.}
\includegraphics[width=1.0\textwidth]{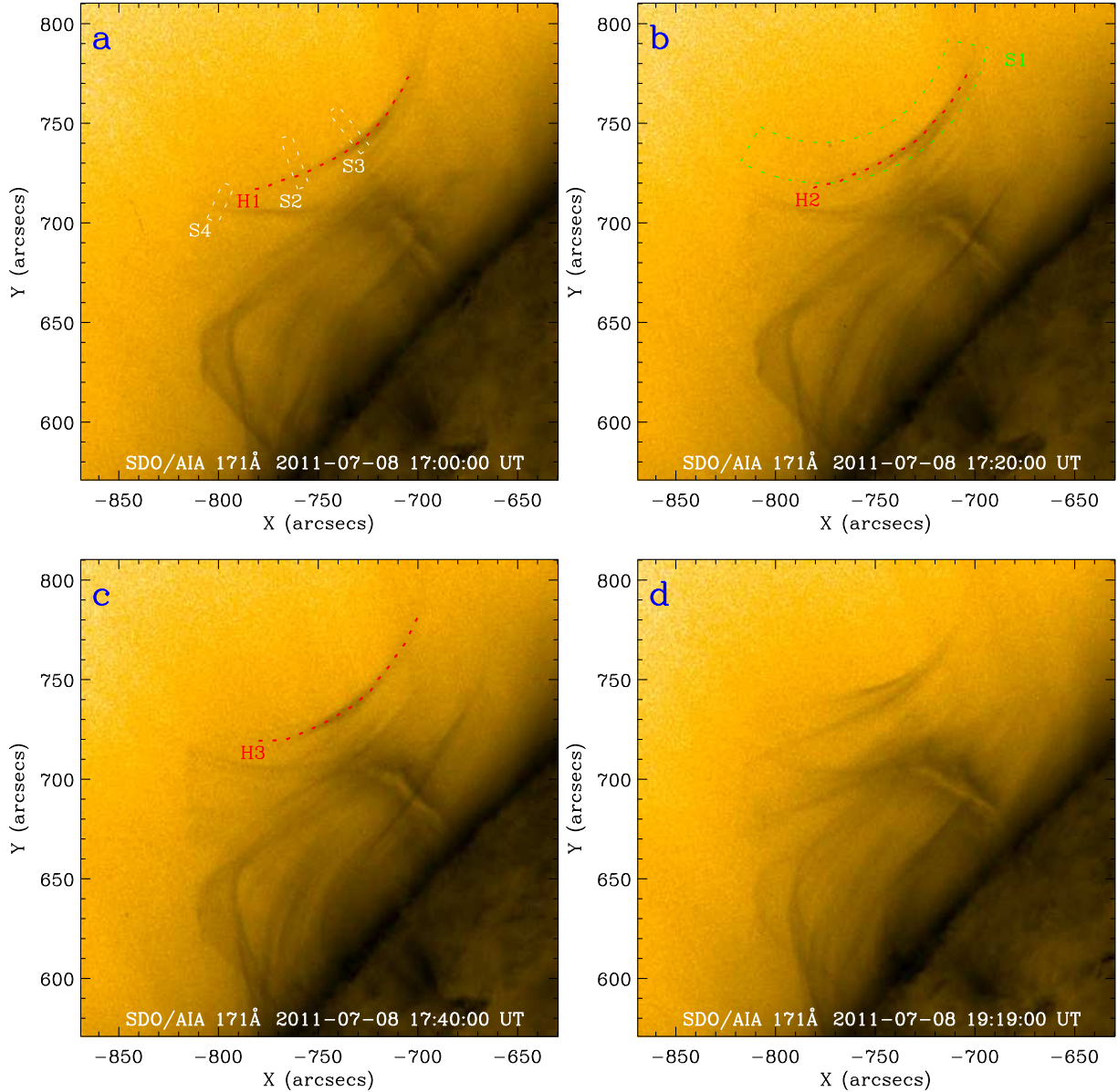}
\caption{Selected horns at different LALO phases in negative
images. The horn profiles are fitted with red dashed circular
arcs. S2-S4 are slices to obtain the distance-time maps across
horns (Figure 4), H1-H3 are to plot the horn flux curves (Figure
5). Their starting points are at the end close to the
labels.}\label{Fig3}
\end{figure}

\begin{figure}
\epsscale{1.}
\includegraphics[width=1.0\textwidth]{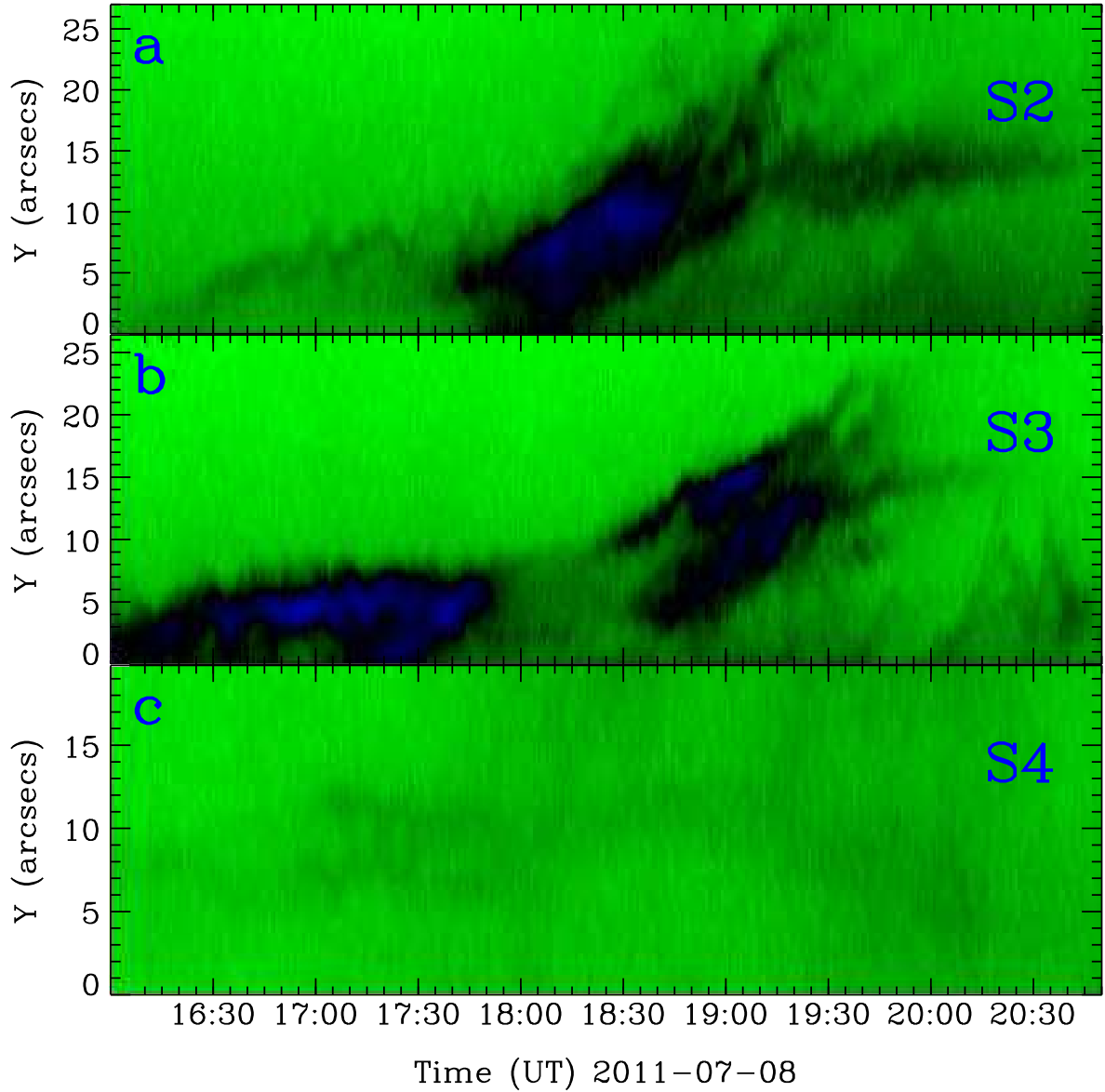}
\caption{Distance-time maps at 171~\AA{} along slices (S2 - S4),
to show the quasi-periodic transverse oscillations of the horns.
An animation of this figure is available. }\label{Fig4}
\end{figure}

\begin{figure}
\epsscale{1.}
\includegraphics[width=1.0\textwidth]{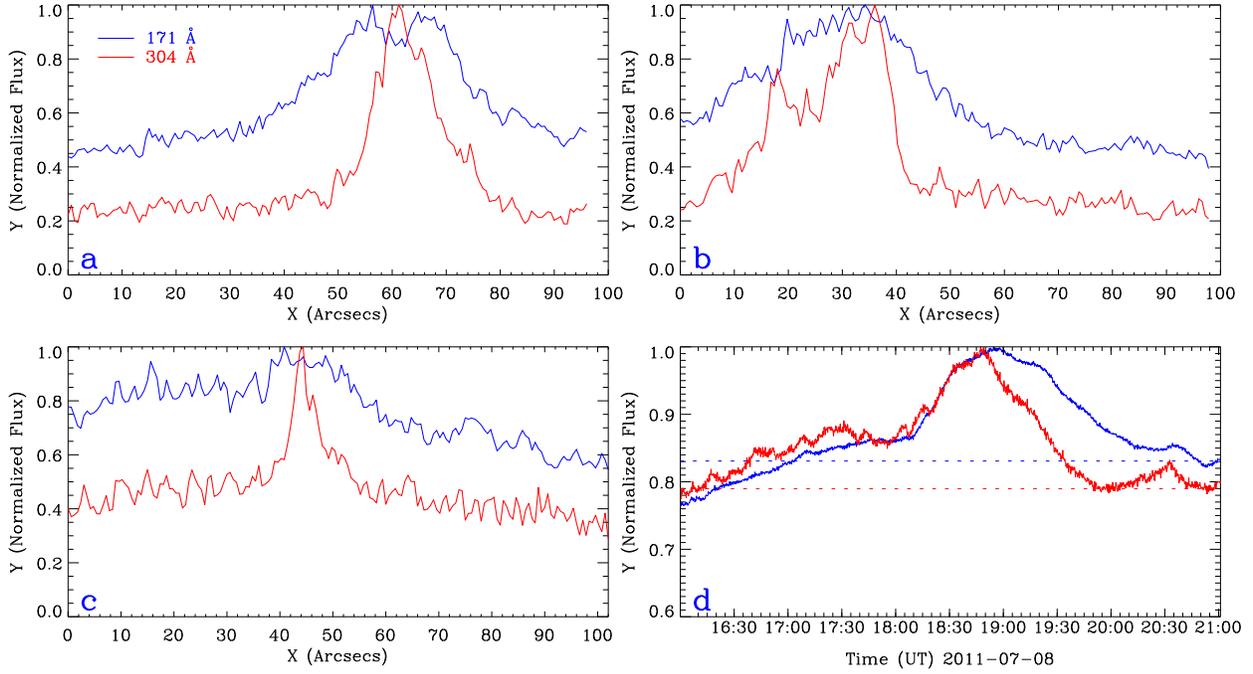}
\caption{Correlation of the AIA 304 (red) and 171 (blue) \AA{}
intensities. (a - c) Normalized intensity variations along the
horns (H1 - H3). (d) The 304 and 171 \AA{} emission intensities
integrated over the window S1, with the background levels
represented by the horizontal dashed lines.}\label{Fig5}
\end{figure}

\end{document}